\documentstyle[emulateapj,apjfonts]{article}

\def\mkfigbox#1#2{
\hbox{ \epsfxsize=#2 \epsfbox{#1} \relax}
}

\def\Sec{${}^{\prime\prime}$\llap{.}}

\def\etal{{\it et~al.\/}}

\def\kpc-1{{kpc$^{-1}$}}
\def\Mpc-1{{Mpc$^{-1}$}}
\def\s-1{{sec$^{-1}$}}
\def\pdeg2{{deg$^{-2}$}}

\def\h0{{H$_0$}}
\def\q0{{$q_0$}}

\def\etal{{et al.}}

\def\ltsima{$\scriptscriptstyle \; \buildrel < \over \sim \;$}
\def\simlt{\lower.3ex\hbox{\ltsima}}

\def\gtsima{$\scriptscriptstyle \; \buildrel > \over \sim \;$}
\def\simgt{\lower.3ex\hbox{\gtsima}}

\def\about{\raise.3ex\hbox{$\scriptscriptstyle \sim $}}

\def\Sec{\hbox{${}^{\prime\prime}$\llap{.}}}

\def\sqr#1#2{{\vcenter{\vbox{\hrule height.#2pt
        \hbox{\vrule width.#2pt height#1pt \kern#1pt
        \vrule width.#2pt}
        \hrule height.#2pt}}}}

\lefthead{Kelson \etal}
\righthead{Absorption Line Strengths from $z$=0 to $z$=0.83}
\submitted{Accepted by Astrophysical Journal Letters 21 March 2001}

\begin{document}

\title{THE EVOLUTION OF BALMER ABSORPTION LINE STRENGTHS\break
IN E/S0 GALAXIES FROM $z$=0 TO $z$=0.83$^1$}

\author{Daniel D. Kelson$^2$, Garth D. Illingworth$^3$, Marijn
Franx$^4$, and Pieter G. van Dokkum$^5$}

\altaffiltext{1}{Based on observations obtained at the W. M. Keck
Observatory, which is operated jointly by the California Institute of
Technology and the University of California.}

\altaffiltext{2}{Department of Terrestrial Magnetism, 5241 Broad
Branch Rd, NW, Washington, DC 20015. Current address: OCIW, 813 Santa
Barbara St, Pasadena, CA 91101; kelson@ociw.edu}

\altaffiltext{3}{UCO/Lick Observatory, Univ. of California, Santa
Cruz, CA 95064}

\altaffiltext{4}{Leiden Observatory, P.O. Box 9513, NL=2300 R.A.
Leiden, The Netherlands}

\altaffiltext{5}{California Institute of Technology, Pasadena, CA}

\begin{abstract}

We present new results from a systematic study of absorption line
strengths of galaxies in clusters approaching redshifts of unity. In
this paper, we specifically compare the strengths of the high-order
Balmer absorption features of H$\gamma$ and H$\delta$ in E/S0s in the
four clusters Abell 2256 ($z$=0.06), CL1358+62 ($z=0.33$), MS2053--04
($z=0.58$), and MS1054--03 ($z=0.83$). By comparing the correlation of
Balmer line strength with velocity dispersion for E/S0s in the three
clusters, we find moderate evolution in the zero-point of the
(H$\delta_A$+H$\gamma_A$)-$\sigma$ relation with redshift. The trend is
consistent with passive evolution of old stellar populations. Under the
assumption that the samples can be compared directly, we use
single-burst stellar population synthesis models to constrain the last
major occurrences of star-formation in the observed E/S0s to be $z_f >
2.5$ (95\% confidence). We have compared the evolution of the Balmer
absorption with the evolution of the $B$-band fundamental plane and find
that simple stellar population models agree very well with the data.
While the best agreement occurs with a low value for $\Omega_m$, the
data provide strong confirmation of the time-evolution in recent stellar
population models.

\end{abstract}


%%%%%%%%%%%%%%%%%%%%%%%%%%%%%%%%%%%%%%%%%%%%%%%%%%%%%%%%%%%%%%%%%%%%%%%%

\section{Introduction}

The star-formation histories of early-type galaxies in distant clusters
have been the focus of many recent studies (e.g.,
\cite{vdf96,bender96,ellis97,kelson97,ziegler,vdfp98,pahre98,kelsonc,z2000}).
While there is broad consensus is that E/S0 galaxies in distant clusters
generally have old stellar populations, these results depend on the
initial mass function (IMF), the cosmology, and the assumption that
morphological evolution can be ignored. Whereas some of these effects
can be taken into account using detailed models (see, e.g.,
\cite{vdfmorph,diaferio}), the interpretation is ultimately limited by
the accuracy of the population synthesis models upon which any
conclusion is based.

These assumptions can now be tested by measuring absorption line
strengths as a function of redshift (time) in early-type cluster
galaxies, The "line strength" results can be compared to the evolution
of colors and $M/L$ ratios, and the correlations between various age
indicators can be compared to predictions of stellar population
synthesis models. Together they can significantly increase our
confidence in the deduced star-formation histories, and in the models
themselves.

In this {\it Letter\/} we present first results on a new study of galaxy
absorption line strengths to redshifts approaching unity, and compare
the measured evolution with the evolution of the galaxy $M/L$ ratios, as
derived from the fundamental plane (see \cite{vdfp98,kelsonc} and
references therein). With this sample, we extend similar work by Bender
\etal\ (1998) by a factor of two in redshift. Here we present the first
results using E/S0 galaxies in the clusters Abell 2256 ($z$=0.06),
CL1358+62 ($z$=0.33), MS2053--04 ($z$=0.58), and MS1054--03 ($z$=0.83),
and discuss the resulting constraints on the star-formation histories,
IMF, cosmological parameters, and the stellar population synthesis
models themselves.

%%%%%%%%%%%%%%%%%%%%%%%%%%%%%%%%%%%%%%%%%%%%%%%%%%%%%%%%%%%%

\section{The Data}

\begin{figure*}[t]
\centerline{ \mkfigbox{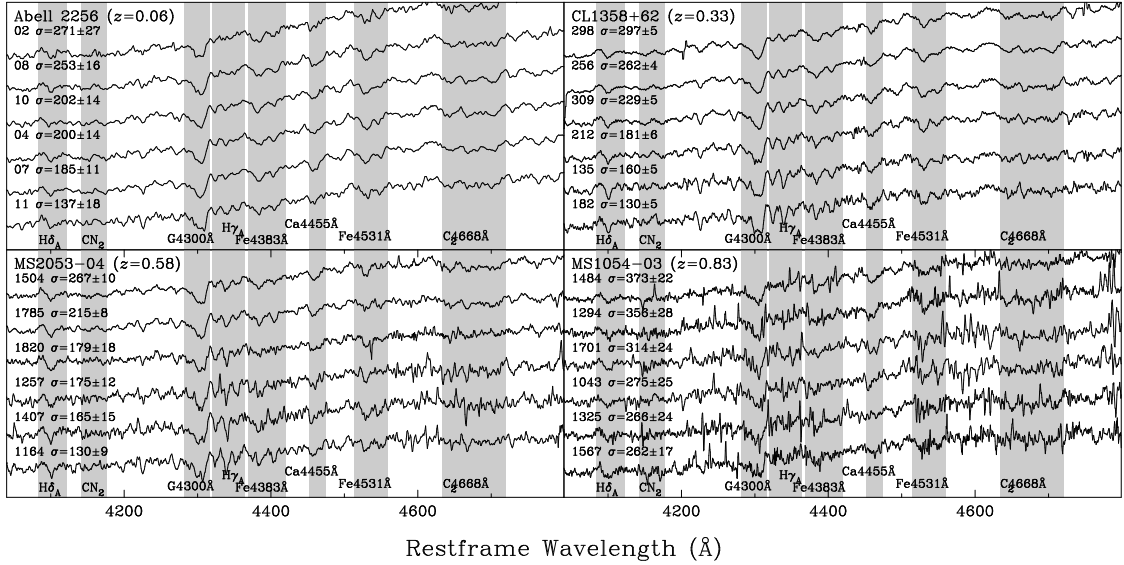}{6.5in}}
\figcaption{Representative spectra for galaxies in Abell 2256
($z$=0.06), CL1358+62 ($z$=0.33), MS2053--04 ($z$=0.58), and MS1054--03
($z$=0.83). The spectra, smoothed using a box width of 3 pixels, have
been shifted to the restframe wavelengths, to illustrate several of the
line strength bandpasses of the Lick/IDS system.
\label{fig:allspec1}}
\end{figure*}

The CL1358+62, MS2053--04, and MS1054--03 spectroscopic data presented
in this paper were obtained at the W.M. Keck Observatory using the Low
Resolution Imaging Spectrograph (\cite{okelris}). The initial sample
selection for the deep spectroscopy was based only on magnitude and
preference for high-resolution spectroscopic follow-up was given to
spectroscopically confirmed members (\cite{fish,tran,vdcm83,tran2}).
Hubble Space Telescope WFPC2 images in F814W were used to determine
morphological classifications (\cite{fabricant,vdcm83,fabric2001}). In
this paper we limit the discussion to the E/S0 galaxies, with more
thorough discussion of the general cluster populations being prepared
for a later paper.

The CL1358+62 spectroscopic data and their reduction were published in
Kelson \etal\ (2000b). The MS2053--04 and MS1054--03 data were similarly
processed. The spectroscopic reductions described in Kelson \etal\
(2000b) provided data of the quality required for measurement of
absorption line strengths, with one additional step for removing
atmospheric H$_2$O absorption at 6800 \AA, 7600 \AA, and 8200 \AA. In
each slit-mask we had included bright blue stars in order to accurately
determine the telluric absorption in each exposure. The galaxy spectra
presented in this paper have $S/N$ ratios ranging from 20-100 per \AA\
(in the continuum). The resolution was typically $\sigma_{inst}\sim
40\hbox{-}60$ km/s. The data for the local comparison sample of 9
early-type galaxies in Abell 2256 were obtained at the KPNO 4m and these
data were processed and analyzed using the same procedures as were used
with the high-$z$ data in an effort to minimize possible systematic
errors. By doing so, we avoid any systematic effects between our distant
samples and those published in the literature (e.g.,
\cite{terlcoma,kunt2000}). Figure \ref{fig:allspec1} shows example
spectra from each of the four clusters.

\begin{figure*}[t]
\centerline{\mkfigbox{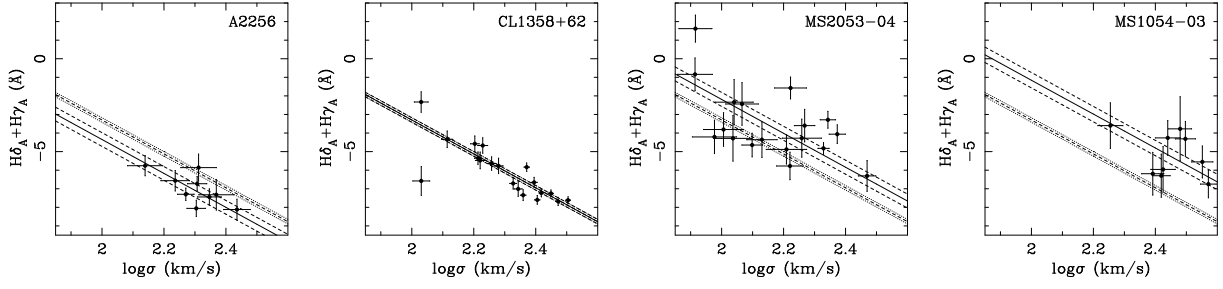}{6.5in}}
\figcaption{The (H$\delta_A$+H$\gamma_A$)-$\sigma$ relations for
early-type galaxies in A2256 ($z$=0.06), CL1358+62 ($z$=0.33),
MS2053--04 ($z$=0.58), and MS1054--03 ($z$=0.83). The best-fitting
linear relation to the CL1358+62 early-type galaxies has a remarkably
low scatter of 13\% in velocity dispersion. While the scatters seen in
A2256 and MS1054--03 are consistent with the observational errors, the
observed scatter in MS2053--04 is \about 50\% larger than expected from
measurement errors alone. The current dataset indicates no statistically
significant differences between ellipticals and S0s, in any of the
clusters. In all four panels, we show the CL1358+62 relation with its
zero-point error (1-$\sigma$). Freezing the slope of the relation, we
indicate the best-fit relations in the other two clusters using the
solid lines, with the zero-point uncertainties indicated by dash lines.
Note the mean shift to larger values of H$\delta_A$+H$\gamma_A$ with
redshift. Increasing values of these indices are the result of stronger
Balmer line absorption at earlier epochs. The high order Balmer indices
evolve significantly with time, providing empirical calibration of the
index's sensitivity to age.
\label{fig:relations}}
\end{figure*}

\begin{figure*}[b]
\centerline{\mkfigbox{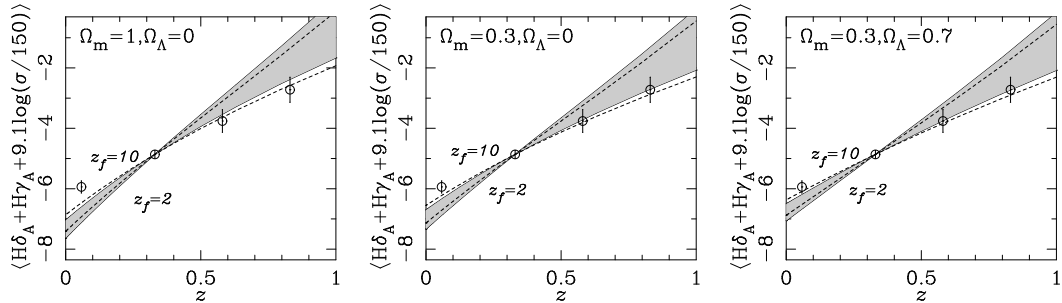}{5.5in}}
\figcaption{The (H$\delta_A$+H$\gamma_A$)-$\sigma$ zero-point evolution
as a function of redshift, with single-burst stellar population models
of Vazdekis \etal\ (1996) (dashed lines) and Worthey (1994; \cite{trager2000})
(shaded). The model curves, shown for formation epochs of $2 \le z_f
\le 10$, are insensitive to the shape of the IMF. Together the data and
models provide lower 95\% confidence limits for the mean epoch of
star-formation of $z_f> 2.9$ and $z_f > 2.4$ using the Worthey and
Vazdekis models, respectively, for $\Omega_m=0.3$, $\Omega_\Lambda=0.7$.
\label{fig:hdahgaz}}
\end{figure*}

%%%%%%%%%%%%%%%%%%%%%%%%%%%%%%%%%%%%%%%%%%%%%%%%%%%%%%%%%%%%

Velocity dispersions were measured using the direct fitting method of
Kelson \etal\ (2000b). The raw velocity dispersions were corrected for
aperture size to values equivalent to a $D=3\Sec 4$ aperture at the
distance of Coma using the prescription of J\o{}rgensen \etal\ (1995).
The corrections were +2.3\%, +4.7\%, +5.6\% and +6.0\% for the galaxies
in A2256, CL1358+62, MS2053--04, and MS1054--03, respectively.

Absorption line strengths were measured using the Trager \etal\ (1998)
definitions for the Lick/IDS indices. For this letter, we use the
indices of the Balmer lines H$\delta$ and H$\gamma$ defined by Worthey
\& Ottavianni (1997), and will discuss the other indices in a later
paper. The raw indices were corrected for both the instrumental
broadening using the Worthey \& Ottavianni (1997) estimates for the
resolution of the Lick/IDS data, and for Doppler broadening.

Accurate estimation of the errors in absorption line strengths requires
knowledge of the variance in the data due to noise (see, e.g.,
\cite{jesus,cardiel}). In work on distant (i.e., highly redshifted)
galaxies, the photon statistics and electronics noise is supplemented by
localized sources of noise, such as sharp residuals from the subtraction
of the bright OH emission lines of the background, and residual fringing
not removed by the flat-fields. Therefore, we estimate the variances
using the residuals between the observed spectra and the high-resolution
model SED from Vazdekis (1999), suitably broadened to match the
instrumental resolution and Doppler broadening of the galaxy spectrum,
which provides the lowest $\chi^2$ (as a function of age and [Fe/H]). We
assume the square of the residuals between the observed and model SEDs
in each index and continuum bandpass accurately reflects the variance
due to noise in the data, and the errors in the absorption line
strengths are derived by propagating these variances. Monte Carlo
simulations have verified that this method provides accurate error
estimates.

Because the metric apertures from which the galaxy spectra were
extracted are systematically larger at high redshift than at low
redshift, the Balmer line strengths needed to be corrected to a
consistent aperture size. Unfortunately no published set of aperture
corrections exist yet for the H$\gamma_A$ and H$\delta_A$ line
strengths. However, one can estimate the corrections by integrating the
published mean major-axis H$\gamma_A$ gradients of the E/S0 galaxies in
Fornax (\cite{kunt1999}), weighted by an $r^{1/4}$-law surface
brightness profile, over a circular aperture. Utilizing the Worthey
(1994) and Vazdekis \etal\ (1996) model prediction that
H$\delta_A+$H$\gamma_A = 1.9\times $H$\gamma_A$, we find that the
required aperture corrections scale with effective aperture diameter as
$\Delta($H$\delta_A+$H$\gamma_A) = (1.78\pm 0.16)\Delta\log D_{ap}$. One
can also scale the Mg$_2$ aperture correction prescription of
J\o{}rgensen \etal\ (1995) by the ratio of the gradients in
(H$\delta_A$+H$\gamma_A$) and Mg$_2$ and recover the same scaling. Using
the above scaling of line strength with aperture size, we correct the
line strengths to a nominal aperture of size $D=1\Sec 23$ at the
distance of CL1358+62 ($z$=0.33). The resulting corrections were
$-0.57^{-0.05}_{+0.05}$\AA, $0.17^{+0.01}_{-0.01}$\AA,
$0.20^{+0.02}_{-0.02}$\AA, for A2256, MS2053--04, and MS1054--03,
respectively.

\begin{figure*}[t]
\centerline{\mkfigbox{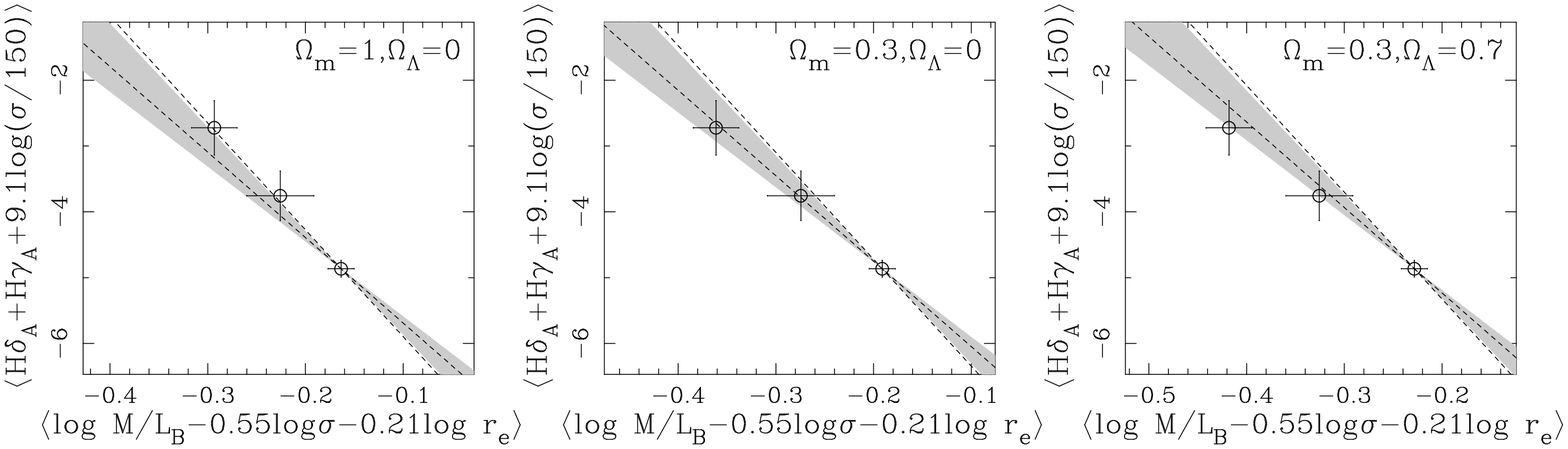}{5.5in}}
\figcaption{The (H$\delta_A$+H$\gamma_A$)-$\sigma$ evolution plotted against
$M/L_B$ evolution, derived from the fundamental plane
(\cite{kelson97,kelsona},b,c, and \cite{vdfp98}). The predicted
correlations of single-burst stellar population models of Vazdekis
\etal\ (1996) (dashed lines) and Worthey (1994) (shaded) are also shown.
The regions are defined by a range of slopes for the initial mass
function $0.85 \le x \le 1.85 $. Note: this projection of
the models only assumes constant metallicity at fixed age and velocity
dispersion, and does not depend on the assumption of co-evality.
\label{fig:hdahgaml}}
\end{figure*}

%%%%%%%%%%%%%%%%%%%%%%%%%%%%%%%%%%%%%%%%%%%%%%%%%%%%%%%%%%%%

\section{Evolution of the Balmer Line Strengths}

In Figure \ref{fig:relations} we show the correlation between
(H$\delta_A$+H$\gamma_A$) (\cite{wortheyhd}) and velocity dispersion for
E/S0 galaxies in A2256, CL1358+62, MS2053--04, and MS1054--03. For the
large sample of early-type galaxies in CL1358+62, there is a
well-defined correlation. This correlation spans a wide range of
velocity dispersion and Balmer line strengths, from positive values
(strong Balmer absorption) to negative values (weak Balmer absorption)
which occur when the mean flux levels in the continuum side-bands are
lower than the mean flux level within the index passbands.

The form of the best-fit linear relation in CL1358+62, shown with its
$\pm1$-$\sigma$ zero-point uncertainties, is $({\rm H}\delta_A+{\rm
H}\gamma_A) = (-9.1\pm 1.0)\log [\sigma/(150\ {\rm km/s})] - (4.9 \pm
0.1)$. The CL1358+62 E/S0s exhibit a remarkably low scatter of 0.50\AA,
equivalent to 13\% in velocity dispersion. When expressed in velocity
dispersion, this scatter is as low as that in the fundamental plane
(\cite{kelsonc}). We have also derived the CL1358+62
H$\gamma_A$-$\sigma$ relation and have compared it to the relation for
Fornax E/S0s (using the published data of \cite{kunt2000}). We find
H$\gamma_A\propto (-3.6\pm 1.4)\log \sigma$ at $z$=0.33 compared to
H$\gamma_A\propto (-4.0\pm 1.5)\log \sigma$ in Fornax, with similarly
low scatter. The lack of strong evolution in the slope of this
correlation suggests that stellar population ages do not vary strongly
along the sequence of early-type galaxies (similar to the conclusions
drawn by \cite{stanford,kelsonc}).

As can be seen in Figure \ref{fig:relations}, the moderately large
observational uncertainties and/or the small sample sizes in the current
dataset prevent an accurate measurement of the slope of the correlation
between Balmer absorption and central velocity dispersion in A2256,
MS2053--04 and MS1054--03.

Given the similarity in the slopes of the Balmer Balmer line-velocity
dispersion relations for Fornax and CL1358+62, we adopt the slope of the
CL1358+62 relation for the remainder of the analysis and assume that
only the zero-point evolves with redshift. We determined the mean
zero-points in the four clusters giving all points uniform weight, using
the bi-weight location estimator (\cite{beers}).

The zero-points for the four clusters, normalized to galaxies with a
dispersion of $\sigma=150$ km/s, are $-5.94\pm 0.20$, $-4.86\pm 0.13$,
$-3.76\pm 0.38$, and $-2.72\pm 0.41$ (in units of \AA). The A2256,
MS2053--04 and MS1054--03 early-type galaxies are offset in zero-point
from CL1358+62 by $\Delta ({\rm H}\delta_A+{\rm H}\gamma_A) = -1.08\pm
0.24$ \AA\ $\Delta ({\rm H}\delta_A+{\rm H}\gamma_A) = 1.10\pm 0.40$
\AA\ and $2.14 \pm 0.43$ \AA, respectively. The offsets (see Figure
\ref{fig:relations}) are significant, 4-, 3- and 5-$\sigma$, and
correspond to larger absorption at earlier epochs (i.e., younger ages),
as expected.

Figure \ref{fig:hdahgaz} shows the evolution of the $({\rm
H}\delta_A+{\rm H}\gamma_A)$ zero-point as a function of redshift.
Assuming the galaxy populations in all three clusters can be compared
directly, we show the predictions of single burst stellar population
models from Worthey (1994; updated in \cite{trager2000}) (shaded) and
Vazdekis \etal\ (1996) (dashed lines). Models with star-formation redshifts
of $2 \le z_f \le 10$ are shown.

Because the model predictions can be simplified to functions of the
form, e.g., H$\gamma_A=A+B\,\delta x+(C\,\delta x +D)\log t+E\log Z + F
(\log Z)^2$, to a high degree of accuracy, we can decouple the
time-evolution of the indices from the metallicity of a given model
(while also rendering the adoption of $H_0=75$ km/s/Mpc unimportant). We
find lower-limits of $z_f > 2.9$ and $z_f > 2.4$ (95\% confidence limit)
using the Worthey and Vazdekis models respectively, for $\Omega_m=0.3,
\Omega_\Lambda=0.7$, similar to conclusions drawn by past work (e.g.,
\cite{vdf96,bender96,kelson97,vdfp98}). As is always the case in this
type of study, the conclusions depend on the assumptions that the
individual cluster samples are co-eval, and that morphological evolution
can be ignored (see, e.g., \cite{vdfmorph}). Furthermore, we also remind
the reader that early-type cluster galaxies may settle into their final
configurations well after their last major epochs of star-formation
leaving their morphological evolution decoupled from their
star-formation histories (e.g., \cite{vdmerge}). 

%%%%%%%%%%%%%%%%%%%%%%%%%%%%%%%%%%%%%%%%%%%%%%%%%%%%%%%%%%%%
\section{Comparison with the Evolution of $M/L$ Ratios}

In past work we have established the evolution of the $M/L_B$ ratio for
the same clusters using the fundamental plane relation
(\cite{kelson97,vdfp98,kelsonc}) and we now compare the mean $M/L_B$
evolution with that of the Balmer lines. The results are shown in Figure
\ref{fig:hdahgaml}. Since the $M/L$ ratios are sensitive to cosmological
parameters, we show the results for several values of $\Omega_m$ and
$\Omega_\Lambda$.

The models agree very well with the observed correlation between the
evolution of the Balmer line strengths and the $M/L_B$ ratios. The slope
of the predicted relation depends on the slope of the IMF through the
$M/L$ ratio evolution (e.g., \cite{tg76}). Therefore, we show the
Worthey (shaded) and Vazdekis (dashed lines) model predictions for a
range of IMF slopes ($0.85 \le x \le 1.85$) about the Salpeter (1955)
value. For standard IMFs and low values of $\Omega_m$, the models are
consistent with our data (and the results of \cite{bender}).

While late-time star-bursts and/or morphological evolution can bias the
observed redshift evolution of the various age indicators (e.g.,
\cite{vdfmorph}), each age-sensitive observable is affected similarly.
Thus, comparisons like that in in Figure \ref{fig:hdahgaml} provide
tests of the population synthesis models which are free from any
assumption of coevality, and are more powerful than simple examinations
of galaxy properties as functions of redshift alone.

By extending our analysis to more spectral indices and galaxies of later
morphological type, we will better constrain the stellar and
morphological histories of cluster galaxies (see, e.g., \cite{kelsonc}).
By doing so, we expect to quantitatively constrain the formation
histories of distant galaxies, and better understand the evolution of
cluster populations as a whole.

%%%%%%%%%%%%%%%%%%%%%%%%%%%%%%%%%%%%%%%%%%%%%%%%%%%%%%%%%%%%

\acknowledgements

We thank I. J\o{}rgensen for assistance in providing the Abell 2256
data, and G. Worthey for kindly making the most recently updated
versions of the Worthey (1994) model programs available. We appreciate
the effort of those at the W.M.Keck observatory who developed and
supported the facility and the instruments that made this program
possible. Lastly we acknowledge the anonymous referee, whose comments
served to greatly improve the paper. Support from STScI grants
GO05989.01-94A, GO05991.01-94A, and AR05798.01-94A, and NSF grant
AST-9529098 is gratefully acknowledged.

%%%%%%%%%%%%%%%%%%%%%%%%%%%%%%%%%%%%%%%%%%%%%%%%%%%%%%%%%%%%


\begin{thebibliography}{reallylongreferencelist}

\bibitem[Beers, Flynn, \& Gebhardt 1990]{beers}Beers, T. C., Flynn,
K., \& Gebhardt, K. 1990, \aj, 100, 32

\bibitem[Bender \etal\ 1998]{bender}Bender, R., Saglia, R.P., Ziegler,
B., Belloni, P., Greggio, L, Hopp, U., \& Bruzual, G. 1998, \apj, 493,
529

\bibitem[Bender, Ziegler, \& Bruzual 1996]{bender96}Bender, R., Ziegler,
B., \& Bruzual, G. 1996, \apjl, 463, L51

\bibitem[Cardiel \etal\ 1998]{cardiel}Cardiel, N., Gorgas, J.,
Cenarro, J, \& Gonz'alez, J.J. 1998, A\&AS, 127, 597

\bibitem[Diaferio \etal\ 2000]{diaferio}Diaferio, A., \etal\ 2000,
\mnras, submitted (astro-ph/0005485)

\bibitem[Ellis \etal\ 1997]{ellis97}Ellis R. S., Smail, I., Dressler,
A., Couch, W.J., Oemler, A., Butcher, H., \& Sharples, R.M. 1997, \apj,
483, 582                                                                         

\bibitem[Fabricant, Franx, \& van Dokkum 2000]{fabricant}Fabricant,
D.G., Franx, M., \& van Dokkum, P.G. 2000, \apj, astro-ph/0003360

\bibitem[Fabricant \etal\ 2001, in prep.]{fabric2001}Fabricant \etal\
2001, in prep.

\bibitem[Fisher \etal\ 1998]{fish}Fisher, D., Fabricant, D.G., Franx, M.,
\& van Dokkum P.G. 1998, \apj, 498, 195

\bibitem[Gonz\'alez 1993]{jesus}Gonz\'alez, J.J. 1993, Ph.D. thesis, Univ.
Calif., Santa Cruz

%\bibitem[Jaffe \etal\ 2000]{boomer}Jaffe, A.H., \etal\ 2000,
%astro-ph/0007333

\bibitem[J\o{}rgensen \etal\ 1995]{jfk95}J\o{}rgensen I., Franx M., \&
Kj\ae{}rgaard P. 1995, \mnras, 276, 1341

\bibitem[J\o{}rgensen 1999]{inger}J\o{}rgensen I. 1999, \mnras, 306,
607

\bibitem[Kelson 1998]{thesis}Kelson, D.D. 1998, Ph.D. thesis, Univ.
Calif., Santa Cruz

\bibitem[Kelson \etal\ 1997]{kelson97}Kelson, D.D., van Dokkum, P.G.,
Franx, M., Illingworth, G.D., \& Fabricant, D.G. 1997, \apjl, 478, L13

\bibitem[Kelson \etal\ 2000a]{kelsona}Kelson, D.D., Illingworth, G.D.,
van Dokkum, P.G., \& Franx, M. 2000a, \apj, 531, 137

\bibitem[Kelson \etal\ 2000b]{kelsonb}Kelson, D.D., Illingworth, G.D.,
van Dokkum, P.G., \& Franx, M. 2000b, \apj, 531, 159

\bibitem[Kelson \etal\ 2000c]{kelsonc}Kelson, D.D. Illingworth, G.D.,
van Dokkum, P.G., \& Franx, M. 2000c, \apj, 531, 184

\bibitem[Kuntschner 1999]{kunt1999}Kuntschner, H. 1999, Ph.D.
dissertation

\bibitem[Kuntschner 2000]{kunt2000}Kuntschner, H. 2000, \mnras, 315,
184

\bibitem[Oke \etal\ 1995]{okelris}Oke, J.B., \etal\ 1995, \pasp, 107, 375

\bibitem[Pahre, de Carvalho, \& Djorgovski 1998]{pahre98}Pahre, M.~A.,
de Carvalho, R.~R., \& Djorgovski, S.~G. 1998, \aj, 116, 1606

\bibitem[Salpeter 1955]{imf}Salpeter, E.E. 1955, \apj, 121, 161

\bibitem[Stanford, Eisenhardt, \& Dickinson 1998]{stanford}Stanford, S.A.,
Eisenhardt, P.R., \& Dickinson, M. 1998, \apj, 492, 461

\bibitem[Terlevich \etal\ 1999]{terlcoma}Terlevich, A.I., Kuntschner,
H., Bower, R.G., Caldwell, N., \& Sharples, R.M., 1999, \mnras, 310, 445

\bibitem[Tinsley \& Gunn 1976]{tg76}Tinsley, B.M. \& Gunn, J.E. 1976, ApJ,
203, 52

\bibitem[Trager \etal\ 1998]{trager}Trager, S.C., Worthey, G., Faber,
S.M., Burstein, D., Gonz\'alez, J.J. 1998, \apjs, 116, 1

\bibitem[Trager \etal\ 2000]{trager2000}Trager, S.C., Faber, S.M.,
Gonz\'alez, J.J., \& Worthey, G. 2000, \apj, submitted

\bibitem[Tran \etal\ 1999]{tran}Tran, K.-V., Kelson, D.D., van Dokkum,
P.G., Franx, G.D., \& Magee, D. 1999, \apj, 522, 39

\bibitem[Tran \etal\ 2001, in prep.]{tran2}Tran \etal\ 2001, in prep.

\bibitem[van Dokkum \& Franx 1996]{vdf96}van Dokkum, P.~G., \& Franx M.
1996, \mnras, 281, 985

\bibitem[van Dokkum \& Franx 2001]{vdfmorph}van Dokkum, P.~G., \& Franx M.
2001, \apj, submitted

\bibitem[van Dokkum \etal\ 1998]{vdfp98}van Dokkum, P.~G., Franx, M.,
Kelson, D. D., \& Illingworth, G. D. 1998, \apjl, 504, L17

\bibitem[van Dokkum \etal\ 1999]{vdmerge}van Dokkum, P.~G., Franx, M.,
Fabricant, D., Kelson, D. D., \& Illingworth, G. D. 1999, \apjl, 520,
L95

\bibitem[van Dokkum \etal\ 2000]{vdcm83}van Dokkum, P.G.,
\etal\ 2000, \apj, 541, 95

\bibitem[Vazdekis 1999]{vaz2a}Vazdekis, A. 1999, \apj, 513, 224

\bibitem[Vazdekis \etal\ 1996]{vaz96}Vazdekis, A., Casuso, E.,
Peletier, R.F., Beckman, J.E. 1996, \apjs, 106, 307

\bibitem[Worthey 1994]{worthey}Worthey, G. 1994, ApJS, 95, 107

\bibitem[Worthey \& Ottaviani 1997]{wortheyhd}Worthey, G. \&
Ottavianni, D.L. 1997, \apjs, 111, 377

\bibitem[Ziegler \& Bender 1997]{ziegler}Ziegler, B. L., \& Bender, R.
1997, MNRAS, 291, 527

\bibitem[Ziegler \etal\ 2000]{z2000}Ziegler, B.L., Bower, R.G., Smail,
I., Davies, R.L., \& Lee, D., 2000, \mnras, submitted

\end{thebibliography}
\end{document}